\documentclass[12pt]{article}
\usepackage{epsfig}
\usepackage{amsfonts}
\usepackage{latexsym}
\usepackage{epic}
\newcommand{\mt}{\mathcal}
\newcommand{\ba}{\begin{array}}
\newcommand{\ea}{\end{array}}
\newcommand{\be}{\begin{equation}}
\newcommand{\ee}{\end{equation}}
\newcommand{\nn}{\nonumber}
\newcommand{\bea}{\begin{eqnarray}}
\newcommand{\ena}{\end{eqnarray}}
\newcommand{\beas}{\begin{eqnarray*}}
\newcommand{\enas}{\end{eqnarray*}}
\newcommand{\mb}{\mbox}

\newcommand{\ac}{\acute}

\begin{document}
\begin{center}
{\LARGE  Simplified tetrahedron equations: Fermionic realization
}
\end{center}
\vspace{2.5 cm}
\begin{center}
{\bf  J.~Ambjorn\footnote{e-mail:{\sl
ambjorn@nbi.dk}}$^{,a,b}$,  ~~Sh.~Khachatryan\footnote{e-mail:{\sl
shah@moon.yerphi.am}}$^{,c}$ and  A.~Sedrakyan\footnote{e-mail:{\sl sedrak@star.yerphi.am}}$^{,a,c}$}\\

\vspace{12pt}

{\it $^a$ Niels Bohr Institute, Blegdamsvej 17, Copenhagen, Denmark\\

\vspace{6pt}

 $^b$ Institute for Theoretical Physics, Utrecht University,\\
Leuvenlann 4, 3584 CE Utrecht, The Netherlands\\

\vspace{6pt}

$^c$ Yerevan Physics Institute, Alikhanian Br. str. 2, Yerevan 36,  Armenia}
\end{center}
\vspace{1.5 cm}

\begin{center}
{\LARGE  Abstract}
\end{center}

The natural generalization of the (two-dimensional) Yang-Baxter
equations to three dimensions is known as the Zamolodchikov's
tetrahedron equations. We consider a simplified version of these
equations which still ensures the commutativity of the transfer
matrices with different spectral parameters and we present a
family of free fermionic solutions.

\newpage
\section{Introduction}

The Algebraic Bethe Ansatz (ABA) is a constructive method for
solving 2d integrable models and it is natural to try to
generalize the method to 3d integrable systems. The key relations
in ABA are the Yang-Baxter equations (YBE), which ensure the
commutativity of two transfer matrices with different spectral
parameters. Zamolodchikov presented a three-dimensional
generalization of the YBE in \cite{Z}. His so-called tetrahedron
equations (ZTE) imply the factorization of the scattering matrices
of more than three one-dimensional objects - strings, thereby
playing the same role for strings as the YBE do for  the
scattering of more than two particles in two-dimensional
integrable field theories \cite{Y,Z1}. ZTE  also generalize
another function of YBE - integrability condition for the lattice
models. They ensure the commutativity of two transfer matrices
with different spectral parameters \cite{Z,Bax}. In the
three-dimensional case the scattering matrices (R-matrices) in the
ZTE play the role of local evolution operators for the states on a
2d lattice (quantum spin model), or of the Boltzmann weights in
the classical statistical models on 3d lattice. There exists a
number of reformulations of ZTE in the context of  integrable 3d
models. In ~\cite{Bax, BB} the 3d conditions for integrability
were formulated with the spin variables assigned to the lattice
sites (interaction around the cube), while the spin variables in
the original Zamolodchikov model were assigned  to the faces of
the 3d lattice. ZTE-equations for spin models where the spin
variables live on the edges of a 3d cubic lattice (the so-called
vertex formulation of ZTE)  can be found in
~\cite{BS,Hieta,Kor,SMS}. In the above listed articles a number of
special solutions have been found and various parameterizations of
ZTE were investigated (see also \cite{KMS,MS,vGPS}). Finally 3d
integrable models have been formulated in terms of free fermions
in \cite{BS3}.

In the first section of this article we once more address the
commutativity of two transfer matrices defined on a 3d lattice in
the vertex formulation and we present a simplification of ZTE
(STE), sufficient for the commutation of transfer matrices with
different spectral parameters and defined on a 2d plane . The
proof that STE implies the commutation of the transfer matrices is
similar the proof in the case of the full ZTE.  We show that in a
particular case the ZTE is formally equivalent to STE. In the last
section we present a description of some solutions of STE which
can be represented in terms of free fermions and we find the
spectrum of the corresponding Hamiltonian.

\section{Commutativity of the transfer matrices.}

 We consider a 3d classical statistical lattice vertex model on a cubic
lattice with periodic boundary conditions. If  $ R_{i,j,k}^{\ac i,\ac j,\ac k} $
is the Boltzmann
weight of a vertex which  joins the links with variables labeled by
$i,j,k,$ and $ \ac i,\ac j,\ac k $, then  the partition function $Z$ of the
model is  defined as the product of the R's over all lattice sites:
\be
 Z=\sum_{\{i,j,k\}}\prod_{all\;vertices}
R_{i,j,k,}^{\ac i,\ac j,\ac k} . \label{z} \ee

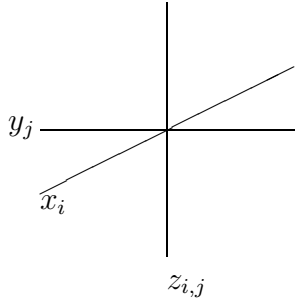
\begin{figure}[t]
\unitlength=12pt
\begin{picture}(100,10)(-10,0)
\put(5,1){\line(0,1){8}}
\put(1,5){\line(1,0){8}}\put(1,3){\line(2,1){8}} \put(0,5){$y_j$}
\put(5,0){$z_{i,j}$} \put(1,2.5){$x_i$}
\end{picture}
\caption{$R_{x_i,y_j,z_{ij}}$ on 3D cubic lattice.} \label{R3f}
\end{figure}

We can rewrite it by defining a monodromy matrix $T$ and the
transfer matrix  $\tau$,  so that :
\be
\ba{ll}
 T=\prod^{' } R_{i,j,k,}^{\ac i,\ac j,\ac k}, \qquad  \tau={\rm tr}\,T,
 \qquad  Z={\rm tr}\, {\tau}^{N_z}.
\ea \ee
 The monodromy matrix $T$ acts on the $ N \times N \times N^2 $ dimensional
spaces (we assume that the lattice is cubic with the equal numbers
of sites in each direction: $N_x=N_y=N_z=N $) and  $\prod^{'}$ denotes the product
only over the horizontal (x y) plane of the 3d lattice.  In the transfer matrix
$\tau$ the trace is taken over the $ N \times N $ dimensional spaces of variables
living on the edge sites of the horizontal (xy) lattice. The variables on
the vertical links (labeled by $z_{ij}$) can be regarded as an elements of  the
$N^2$ dimensional quantum space,  while  the horizontal ones form a
$N \times N$ dimensional so-called auxiliary space, labeled by $x_i, z_j$ (see
Fig.\ \ref{R3f} for an illustration of the notation $R_{x_i y_j z_{ij}}$).

In order to find a condition which ensures that two transfer
matrices $\tau(u)$ and $\tau(v)$ with different spectral
parameters, $u$ and $v$, commute,  we are going to define an
extension of the YBE known from 2d models to the 3d models.  In
analogy with the method used for the YBE we will define two
intertwiner $R(u,v)$ matrices which will ensure the commutativity
of the transfer matrices. At this point it should be mentioned
that when we talk about spectral parameters in higher than two
dimensions they should not be thought of as single real variable
like in 2d, but as a set of continuous variables or angles on
which the matrices $R_{ijk}$ depend.

The transfer matrix
constructed from  the $R_{ijk}$-matrices can be represented as
\bea
\tau(u)=\sum_{\{xy\}}\prod_{i=1}^{N}\prod_{j=1}^{N}
R_{x_i,y_j,z_{ij}}(u)
\ena
Let us denote the variables of the auxiliary spaces by $x_i$ and
$y_j$ in the first transfer matrix $\tau(u)$ and by $a_i, b_j$ in the second
transfer matrix $\tau(v)$. As above the $N\times N$ states on the vertical links  are denoted
by $z_{ij}$ (see Fig.\ \ref{R3f}). Then the product of two transfer
matrices is
\bea
\label{X}
\tau(u)\tau(v)=\qquad  \qquad\qquad  \qquad \qquad \qquad \qquad \qquad\\
=\sum_{\{x,y; a,b\}}\left(R_{x_1,y_1,z_{11}}(u)\cdots
R_{x_1,y_N,z_{1N}}(u)R_{x_2,y_1,z_{21}}(u)
R_{x_2,y_N,z_{2N}}(u)\cdots R_{x_N,y_N,z_{NN}}(u)\right)\nn \\
\times \left(R_{a_1,b_1,z_{11}}(v)\cdots
R_{a_1,b_N,z_{N1}}(v)R_{a_2,b_1,z_{21}}(v) \cdots
R_{a_2,b_N,z_{2N}}(v)\cdots R_{a_N,b_N,z_{NN}}(v)\right). \nn \ena
 We expect, that there is an intertwiner operator
$R(u,v)$ ensuring the commutativity of the transfer matrices with
different spectral parameters: \be R(u,v)T(u)T(v)=T(v)T(u)R(u,v),
\ee \be \tau(u)\tau(v) ={\rm tr}\;(T(u)T(v))={\rm
tr}\;(R(u,v)T(v)T(u)R^{-1}(u,v)).\label{TT} \ee $R$ matrix acts on
the space labeled by the variables $a_i,b_i,x_i,y_i$. Let us
define the intertwiner R matrix as the product of the local
intertwiners: \be R_{a,b;x,y}(u,v)=R_{x_N,a_N}(u,v)\cdots
R_{x_1,a_1}(u,v)R_{y_1,b_1}(u,v)\cdots
 R_{y_N,b_N}(u,v).
\label{int}
\ee
By use of a so-called railway argument one can  prove
the commutativity condition (\ref{TT}):
 \bea
 \label{com}
&& {\rm tr}\; \{T(u)T(v)\}={\rm tr}\; \{R_{a,b;x,y}(u,v)T(u)T(v)R^{-1}_{a,b;x,y}\}= \\
&=&\sum_{\{a,b,x,y\}} \{R_{x_N,a_N}(u,v)\cdots R_{y_N,b_N}(u,v)
\prod_{i=1}^{N}\prod_{j=1}^{N}\left(
R_{x_i,y_j,z_{ij}}(u)R_{a_i,y_j,z_{ij}}(v)\right)\nn  \nn \\
&\times& R^{-1}_{y_N,b_N}(u,v)\cdots R^{-1}_{x_N,a_N}(u,v)\}=
\sum_{\{a,b,x,y\}}\prod_{i=N}^{2}R_{x_i,a_i}(u,v)R_{y_i,b_i} \{
R_{x_1,a_1}(u,v)R_{y_1,b_1}(u,v)\nn  \\
&\times& R_{x_1,y_1,z_{11}}(u)R_{a_1,b_1,z_{11}}(v)\} \prod_{i=2}^{N}
R_{y_i,b_i}(u,v)\left(\prod_{j=2}^{N}R_{x_1,y_j,z_{1j}}(u)R_{a_1,y_j,z_{1j}}(v)\right.  \nn \\
&\times& \left.
\prod_{i=2}^{N}\prod_{j=1}^{N}R_{x_i,y_j,z_{ij}}(u)R_{a_i,y_j,z_{ij}}(v)\right)
\prod_{i=N}^{1}R^{-1}_{y_i,b_i}(u,v)\prod_{i=1}^{N}R^{-1}_{x_i,a_i}(u,v).   \nn
\ena
While the ordering  of the $R_{x_ia_j}$ and  $R_{b_iy_j}$ operators
is not important since they commute, we shall nevertheless treat  them as
non-commuting as long as possible (see later).

If  the following equations for local R-matrices are fulfilled for every $i,j=1,...,N$:
\be \ba{ll}
R_{x_i,a_i}(u,v)&R_{y_j,b_j}(u,v)R_{x_i,y_j,z_{jj}}(u)R_{a_i,b_j,z_{ij}}(v)=\\
&=R_{a_i,b_j,z_{ij}}(v)R_{x_i,y_j,z_{ij}}(u)R_{y_j,b_j}(u,v)R_{x_i,a_i}(u,v).
\ea \label{sted}
\ee
then by taking them into account in the figure brackets of the equation
(\ref{com}) and successively continuing that procedure along of all plane
we will get
\bea\nn \tau(u)\tau(v)&=&\sum_{a,b;x,y} { }
\prod_{i=N}^{2}R_{x_i,a_i}(u,v)
\{R_{a_1,b_1,z_{11}}(v)R_{x_1,y_1,z_{11}}(u)
R_{y_1,b_1}(u,v)R_{x_1,a_1}(u,v)\}\\\nn
&\times&R_{y_2,b_2}(u,v)R_{x_1,y_2,z_{12}}(u)R_{a_1,b_2,z_{12}}(v)
\left(\prod_{i=3}^{N}R_{y_i,b_i}(u,v)R_{x_1,y_i,z_{1i}}(u)R_{a_1,b_i,z_{1i}}(v)\right.\\\nn
&\times&\left.\prod_{i=2}^{N}\prod_{j=1}^{N}R_{x_i,y_j,z_{ij}}(u)R_{a_i,y_j,z_{ij}}(v)\right)
\prod_{i=N}^{1}R^{-1}_{y_i,b_i}(u,v)\prod_{i=1}^{N}R^{-1}_{x_i,a_i}(u,v)=\nn \\
=\cdots& =&\sum \prod_{i=N}^{2}R_{x_i,a_i}
\left(\prod_{j=1}^{N}R_{a_1,b_j,z_{1j}}(v)R_{x_1,y_j,z_{1j}}(u)\right)
\prod_{i=1}^{N}R_{y_i,b_i}(u,v) R_{x_1,a_1}(u,v)\nn \\
&\times&\left(\prod_{i=2}^{N}\prod_{j=1}^{N}R_{x_i,y_j,z_{ij}}(u)R_{a_i,y_j,z_{ij}}(v)\right)
\prod_{i=N}^{1}R^{-1}_{y_i,b_i}(u,v)\prod_{i=1}^{N}R^{-1}_{x_i,a_i}(u,v)=\nn \\
&=&\sum { }
\left(\prod_{j=1}^{N}R_{a_1,b_j,z_{1j}}(v)R_{x_1,y_j,z_{1j}}(u)\right)
\prod_{i=N}^{3}R_{x_i,a_i}(u,v)
\left(R_{a_2,b_1,z_{21}}(v)R_{x_2,y_1,z_{21}}(u)\right)\nn \\
&\times&R_{y_1,b_1}(u,v)
R_{x_2,y_2(u,v)}\prod_{i=2}^{N}R_{y_i,b_i}(u,v)\left(
\prod_{j=2}^{N}R_{x_2,y_j,z_{2j}}(u)R_{a_2,b_j,z_{2j}}(v)\right.\nn \\
&\times&\left.
\prod_{i=3}^{N}\prod_{j=1}^{N}R_{x_i,y_j,z_{ij}}(u)R_{a_i,y_j,z_{ij}}(v)\right)
R_{x_1,a_1}(u,v)
\prod_{i=N}^{1}R^{-1}_{y_i,b_i}(u,v)\prod_{i=1}^{N}R^{-1}_{x_i,a_i}(u,v)\nn \\
=\cdots &=&\sum
\left(\prod_{i=1}^{N}\prod_{j=1}^{N}R_{a_i,b_j,z_{ij}}(v)R_{x_i,y_j,z_{ij}}(u)\right)
\prod_{i=1}^{N}R_{y_i,b_i}(u,v)\prod_{i=N}^{1}R_{x_i,a_i}(u,v) \nn \\
&\times&\prod_{i=N}^{1}R^{-1}_{y_i,b_i}(u,v)\prod_{i=1}^{N}R^{-1}_{x_i,a_i}(u,v).
 \label{sem}\ena

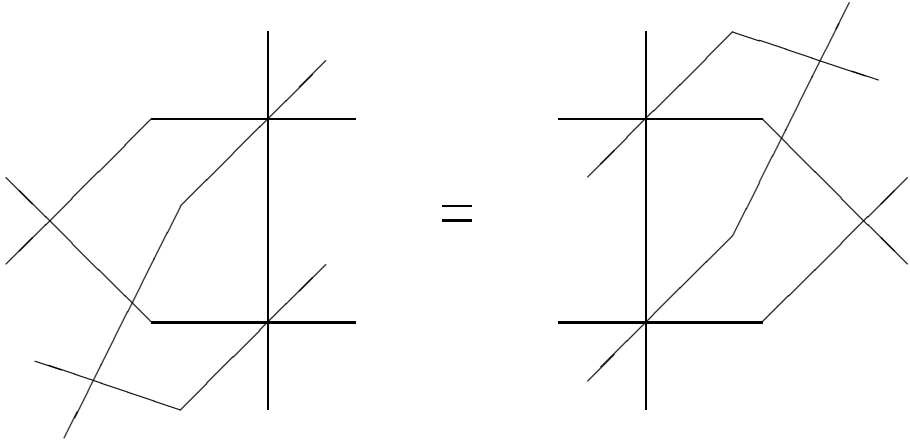
\begin{figure}[h]
\unitlength=11pt
\begin{picture}(100,20)(0,0)
\put(10,2){\line(0,1){13}}
\put(6,5){\line(1,0){7}}\put(7,2){\line(1,1){5}}
\put(6,12){\line(1,0){7}}\put(7,9){\line(1,1){5}}\put(6,12){\line(-1,-1){5}}
\put(6,5){\line(-1,1){5}}
\put(7,9){\line(-1,-2){4}}\put(7,2){\line(-3,1){5}}
\multiput(16,8.5)(0,0.5){2}{\line(1,0){1}}
\put(23,2){\line(0,1){13}}
\put(20,5){\line(1,0){7}}\put(26,8){\line(-1,-1){5}}
\put(20,12){\line(1,0){7}}\put(26,15){\line(-1,-1){5}}\put(26,8){\line(1,2){4}}\put(26,15){\line(3,-1){5}}
\put(27,5){\line(1,1){5}}\put(27,12){\line(1,-1){5}}\end{picture}
\caption{Graphical representation of STE} \label{STEEE}
\end{figure}

As $R_{x_i,a_i}$ and $R_{y_i,b_i}$ commute it is seen from the
last equation in (\ref{sem}) that \be \label{1}
\tau(u)\tau(v)=\tau(v)\tau(u). \ee Equations (\ref{sted}) can be
represented graphically as in (\ref{STEEE}).  We call them
Semi-tetrahedron equations (STE) (referring to the known equations
of Zamolodchikov).

If we force the intertwiner matrices to have a structure similar
to the $R_{ijk}$  (the Boltzmann weights), we must consider the
$R_{ab}^{\ac a\ac b}$ matrix elements as operators acting on an
additional (auxiliary)  space (the \textit{same} space for all
local intertwiners): $R_{c,x_i,a_i}, R_{c,y_i,b_i}$, and then take
the trace in the transfer matrices over that space. In this case
the intertwiner in the equation (\ref{int}) will be
 \be
R_{a,b;x,y;c}(u,v)=R_{c,x_N,y_N}(u,v)\cdots
R_{c,x_1,y_1}(u,v)R_{c,a_1,b_1}(u,v)\cdots
 R_{c,a_N,b_N}(u,v).
\label{int1}
\ee
and  the commutativity conditions of transfer matrices (\ref{sted}) are defined by
 Zamolodchikov's tetrahedron equations:
\be
\ba{l}
R_{c,x_i,a_i}(u,v)R_{c,y_i,b_i}(u,v)R_{x_i,y_i,z_{ii}}(u)R_{a_i,b_i,z_{ii}}(v)=\\
\hspace{2cm}R_{a_i,b_i,z_{ii}}(v)R_{x_i,y_i,z_{ii}}(u)R_{c,y_i,b_i}(u,v)R_{c,x_i,a_i}(u,v).
\ea \label{tetr}
\ee
Up to the last step, all equations in (\ref{sem}) are valid in this general case
as well, but the commutativity of  intertwiners $R_{x_i,a_i}$ and $R_{y_i,b_i}$ is broken
for general $R_{c,x_i,a_i}$ and
$R_{c,y_i,b_i}$ and we have to make a more detailed analysis.
First we should take trace in the
space of new auxiliary variable $c$
 \be
{\rm tr}_c \{ R_{y,b,c}\otimes R_{x,a,c}\}\{T_{x,y}\otimes
T_{a,b}\} = \{T_{a,b}\otimes T_{x,y}\}\,{\rm
tr}_c\{R_{x,a,c}\otimes R_{y,b,c}\}, \ee \be
R_{x,a,c}=\prod_{i}R_{c,x_i,y_i},\quad  R_{y,b,c}=\prod_{i}
R_{c,x_i,a_i}. \ee As the variables $x,a$ and $y,b$ are
independent, it is possible to change the positions of the
R-matrices under the trace \be {\rm tr}_c \{ R_{x,a,c}\otimes
R_{y,b,c}\} = {\rm tr}_c \{ R_{y,b,c}\otimes R_{x,a,c}\} \ee and
following the by now familiar steps we arrive at the desired
result
 \be\nn \ba{l} {\rm tr}_c \{
R_{y,b,c}\otimes R_{x,a,c}\}\{T_{x,y}\otimes T_{a,b}\} ({\rm tr}_c
\{ R_{x,a,c}\otimes R_{y,b,c}\})^{-1}=\\
= {\rm tr}_c \{ R_{y,b,c}\otimes R_{x,a,c}\}\{T_{x,y}\otimes
T_{a,b}\} ({\rm tr}_c \{ R_{y,b,c}\otimes R_{x,a,c}\})^{-1}, \ea
\ee \be\nn \ba{c} {\rm tr}_{a,b,x,y}\{T_{x,y}(u)\otimes
T_{a,b}(v)\}=\\= {\rm tr}_{a,b,x,y}({\rm tr}_c \{ R_{y,b,c}\otimes
R_{x,a,c}\}\{T_{x,y}(u)\otimes T_{a,b}(v)\} ({\rm tr}_c \{
R_{y,b,c}\otimes R_{x,a,c}\})^{-1})=\\={\rm
tr}_{a,b,x,y}\{T_{a,b}(u)\otimes T_{x,y}(v)\}. \ea \ee

 In the Figure 2 we present graphically the semi-tetrahedron and the tetrahedron equations,
respectively.  It is seen that the semi-tetrahedron equations can be decomposed in two
connected triangle equations.
\unitlength=2.6mm
\newsavebox{\blok}
\sbox{\blok}{
\begin{picture}(22,21)
\put(7,3){\line(3,2){11}} \put(13,3){\line(-3,2){11}}
\put(10,2){\line(0,1){18}} \put(2.5,7){\line(3,4){9}}
\put(17.5,7){\line(-3,4){9}} \put(6.7,3.7){$x$}
\put(10.2,2.5){$z$} \put(12.7,3.7){$y$} \put(8.5,19){$a$}
\put(11.7,19){$b$} \put(13.5,6.2){$x$} \put(18,10.8){$x$}
\put(6,6.2){$y$} \put(2,10.8){$y$} \put(10.2,12){$z$}
\put(10.2,18.8){$z$}
\multiput(21,8)(0,0.5){2}{\line(1,0){1}} \put(13.8,13.5){$a$}
\put(18,7.2){$a$} \put(6.3,13.5){$b$} \put(1.6,7){$b$}
\end{picture}}

\newsavebox{\blokone}
\sbox{\blokone}{\begin{picture}(22,21) \put(1,11){\line(3,2){11}}
\put(19,11){\line(-3,2){11}} \put(10,2){\line(0,1){18}}
\put(8.5,3){\line(3,4){9}} \put(11.5,3){\line(-3,4){9}}
\put(0.7,11.7){$x$} \put(10.2,2.5){$z$} \put(18.7,11.7){$y$}
\put(2.5,15){$a$} \put(17.6,15){$b$} \put(7.5,14.2){$ x$}
\put(12,18.8){$x$} \put(12,14.2){$y$} \put(8,18.8){$y$}
\put(10.2,11){$z$} \put(10.2,18.8){$z$} \put(7.8,9.5){$a$}
\put(12,3.2){$a$} \put(12.3,9.5){$b$} \put(7.8,3){$b$}
\end{picture}}

\newsavebox{\bloktwo}
\sbox{\bloktwo}{
\begin{picture}(22,21)
\put(7,3){\line(3,2){11}} \put(13,3){\line(-3,2){11}}
\put(10,2){\line(0,1){18}} \put(2.5,7){\line(3,4){9}}
\put(17.5,7){\line(-3,4){9}} \put(6.7,3.7){$x$}
\put(10.2,2.5){$z$} \put(12.7,3.7){$y$} \put(8.5,19){$a$}
\put(11.7,19){$b$} \put(13.5,6.2){$ x$} \put(18,10.8){$x$}
\put(6,6.2){$y$} \put(2,10.8){$y$} \put(10.2,12){$z$}
\put(10.2,18.8){$z$}
 \multiput(21,8)(0,0.5){2}{\line(1,0){1}}
\put(13.8,13.5){$a$} \put(18,7.2){$a$} \put(6.3,13.5){$b$}
\put(1.6,7){$b$} \put(1,9){\line(1,0){18}} \put(1,9.2){$c$}
\put(9,9.2){$c$} \put(19,9.2){$c$}
\end{picture}}

\newsavebox{\blokthree}
\sbox{\blokthree}{\begin{picture}(22,21)
\put(1,11){\line(3,2){11}} \put(19,11){\line(-3,2){11}}
\put(10,2){\line(0,1){18}} \put(8.5,3){\line(3,4){9}}
\put(11.5,3){\line(-3,4){9}} \put(0.7,11.7){$x$}
\put(10.2,2.5){$z$} \put(18.7,11.7){$y$} \put(2.5,15){$a$}
\put(17.6,15){$b$} \put(7.5,14.2){$ x$} \put(12,18.8){$x$}
\put(12,14.2){$y$} \put(8,18.8){$y$} \put(10.2,11){$z$}
\put(10.2,18.8){$z$} \put(7.8,9.5){$a$} \put(12,3.2){$a$}
\put(12.3,9.5){$b$} \put(7.8,3){$b$} \put(1,13){\line(1,0){18}}
\put(1,13.2){$c$} \put(9,13.2){$c$} \put(19,13.2){$c$}
\end{picture}}
\vspace{0.5cm}
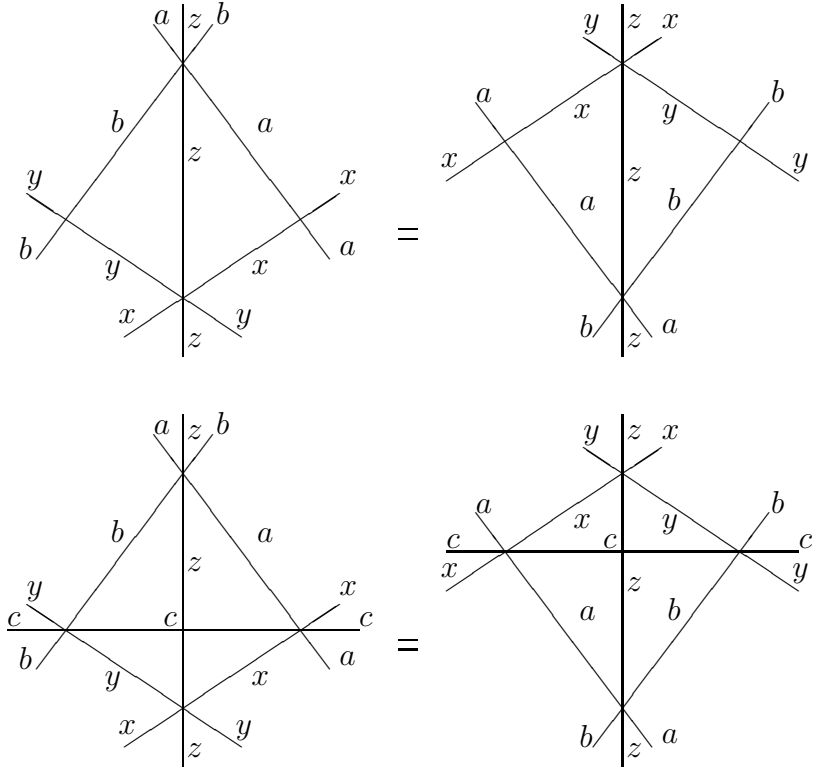
\begin{figure}[h]
\begin{picture}(46,44)(-6.5,0)
\put(0,22){\usebox{\blok}} \put(23,22){\usebox{\blokone}}
\put(0,1){\usebox{\bloktwo}} \put(23,1){\usebox{\blokthree}}
\end{picture}
\vspace{-0.5cm} \caption{Graphical representation of the semi-tetrahedron
equations (the upper pair) and the tetrahedron equation (the lower pair).}
\end{figure}
The semi-tetrahedron equations  are cases of the tetrahedron
equations. It is seen that if \be
tr_c(R_{axc}R_{byc})=R_{xa}R_{yb} \ee the tetrahedron equations
reduce to the  semi-tetrahedron equations. Thus the structure of
the  STE is somewhat simpler than the structure of the ZTE and
they consist of fewer equations than ZTE.

\section{Solutions of the STE using free fermionic representations.}
\noindent

We can present rather simple solutions of the STE  by
using a free fermionic realization of  R-operator
when  the representation
spaces where it acts are two-dimensional,
i.e.\ the spaces of spin 1/2 states on the links of the
3d lattice.

In the  two-dimensional integrable
models considered in \cite{ShA, AShS0, AShS} the three-particle
$R_{ijk}$ matrices were represented  in terms of fermionic
operators:
\bea
 \textbf{R}_{123}:|i_1\rangle\otimes|i_2\rangle\otimes|i_3\rangle \rightarrow
 |j_1\rangle\otimes|j_2\rangle\otimes|j_3\rangle, \qquad  \qquad \qquad \qquad \qquad \nn \\
 \textbf{R}_{123}=(-1)^{(p(i_1)+p(i_2))
 (p(i_3)+p(j_3))+p(i_1)(p(i_2)+p(j_2))}(\textbf{R}_{123})^{j_1 j_2 j_3}_{i_1 i_2 i_3}
 (|j_1\rangle \langle i_1|)
 (|j_2\rangle \langle i_2|)
 (|j_3\rangle \langle i_3|),\nn \\
 {(i,j)}_{(1,2,3)}=0,1, \qquad \qquad \qquad \qquad \qquad \qquad \qquad \qquad
\label{R3}
\ena
with the fermionic creation and annihilation operators
for each representation space defined as
\bea
(|0\rangle \langle
0|)_i=1-n_i,\quad (|1\rangle \langle 1|)_i=n_i,\quad (|0\rangle
\langle 1|)_i=c_i,\quad (|1\rangle \langle 0|)_i=c^+_i,\quad
n_i=c^+_i c_i .
\ena
Here $p(\alpha)=\alpha$ are the parities of
the states, due to the graded character of fermionic states
\cite{AShS}.  The R-matrices we will use here differ from the above
matrices by a permutation operator
\bea
P_{13}|1\rangle\otimes|2\rangle\otimes|3\rangle=|3\rangle\otimes|2\rangle\otimes|1\rangle,
\qquad \qquad \qquad \qquad \qquad \qquad \\\label{matrixR}
R_{123}=P_{13}\textbf{R}_{123}=(-1)^{p(i_1)
 (p(j_2)+p(j_3))+p(i_2)p(j_3)}(R_{123})^{j_1 j_2 j_3}_{i_1 i_2 i_3}
 (|j_1\rangle \langle i_1|)
 (|j_2\rangle \langle i_2|)
 (|j_3\rangle \langle i_3|).
 \ena
Similarly a  representation for the two-particle R-matrix can be defined as
\bea
\label{matrixR1}R_{12}:|1\rangle|2\rangle\rightarrow|2\rangle|1\rangle,\quad
R_{12}=(-1)^{p(i_1)
 p(j_2)}(R_{12})^{j_1 j_2 }_{i_1 i_2}
 (|j_1\rangle \langle i_1|)
 (|j_2\rangle \langle i_2|).
  \ena
A representation of the  three-particle $R$ operators with
$R_{ijk}$-matrices which cannot be reduced to products of two-particle
R-matrices was found in \cite{AShS}:
\bea
\label{a}R_{123}=:e^{(a_{ij}-\delta_{ij})c^+_{i}c_{j}}:,\quad
i,j=1,2,3,\quad a_{ii}=0,
\ena
where $:():$ means normal ordering.

 In general, when $a_{ii}\neq 0$, the representation above has 9 independent
 parameters $a_{ij}$
(not counting the normalization
 parameter).  On the other hand the
 $R_{123}$ operator in the so-called
 "free-fermionic" representation, $:e^{\sum_{i,j}^3(a_{ij}-\delta_{ij})c^+_{i}c_{j}}:$,
 (called so since the fermions enter
 quadratically) conserves the number of the fermions and has 20 matrix elements.
 So in this case there are 10 constrains between the matrix
 elements in (\ref{matrixR}).

 Similarly, the $R_{12}$ operator $:e^{\sum_{i,j}^2(a_{ij}-\delta_{ij})c^+_{i}c_{j}}:$
 has 4 independent
 parameters $a_{ij}$ and one normalization parameter, so one constraint, known from the
 $XX$ model is imposed on the six matrix elements (\ref{matrixR1}):
 \bea
 R_{00}^{00}R_{11}^{11}+R_{01}^{10}R_{10}^{01}=R_{01}^{01}R_{10}^{10}.
 \ena
 Let us write down all ten relations between the $R_{123}$ matrix
 elements, having the above "free-fermionic" representation.
\bea
R_{010}^{001}R_{100}^{100}+R_{110}^{101}R_{000}^{000}=R_{010}^{100}R_{100}^{001},\qquad
R_{001}^{010}R_{100}^{100}+R_{101}^{110}R_{000}^{000}=R_{001}^{100}R_{100}^{010},\nn\\
R_{110}^{011}R_{000}^{000}+R_{100}^{010}R_{010}^{001}=R_{100}^{001}R_{010}^{010},\qquad
R_{011}^{110}R_{000}^{000}+R_{001}^{010}R_{010}^{100}=R_{001}^{100}R_{010}^{010},\nn\\
R_{101}^{011}R_{000}^{000}+R_{100}^{010}R_{001}^{001}=R_{100}^{001}R_{001}^{010},\qquad
R_{010}^{100}R_{001}^{001}+R_{011}^{101}R_{000}^{000}=R_{010}^{001}R_{001}^{100},\nn\\
R_{110}^{110}R_{000}^{000}+R_{100}^{100}R_{010}^{010}=R_{100}^{010}R_{010}^{100},\qquad
R_{011}^{011}R_{000}^{000}+R_{001}^{001}R_{010}^{010}=R_{010}^{001}R_{001}^{010},\nn\\
R_{101}^{101}R_{000}^{000}+R_{001}^{001}R_{100}^{100}=R_{001}^{100}R_{100}^{001},\qquad \hspace{2.5cm} \nn\\
R_{000}^{000}R_{111}^{111}=det\left( \ba{ccc}
R_{100}^{100}&R_{100}^{010}&R_{100}^{001}\\
R_{010}^{100}&R_{010}^{010}&R_{010}^{001}\\
R_{001}^{100}&R_{001}^{010}&R_{001}^{001}
\ea \right). \qquad \qquad \qquad \qquad
 \ena
In terms of $a_{ij}$ parameters  equations (\ref{sted})  contain
25 constraints - the independent ones from the $2^5\times 2^5$
equations for matrix elements. Substituting (\ref{matrixR}) and
(\ref{matrixR1}) in the STE equations (\ref{sted}) we find
 \bea
\sum_{j_\alpha}R_{i_x i_a}^{j_x j_a}(u,v)R_{i_y i_b}^{j_y
j_b}(u,v)R_{j_x j_y i_z}^{k_x k_y j_z}(u) R_{j_a j_b j_z}^{k_a k_b
k_z}(v)(-1)^{p(j_x)p(j_b)+p(k_y)p(k_a)}=\nn
\\
\sum_{j_\alpha}R_{i_a i_b i_z}^{j_a j_b j_z}(v)R_{i_x i_y
j_z}^{j_x j_y k_z}(u) R_{j_y j_b}^{k_y k_z}(u,v)R_{j_x j_a}^{k_x
k_a}(u,v)(-1)^{p(j_x)p(j_b)+p(i_y)p(i_a)}, \quad
i_x,i_y,...=0,1.\label{stede}\ena

Note that when dealing with fermionic spaces as here, in the
matrix representation of the vertex ZTE (\ref{tetr}), some
additional signs will appear in the equations (see the Appendix
for details), compared to the usual form of these equations
\cite{BS}. These signs are reflecting fermionic (graded) nature of
our redefinitions of  ZTE.

 Looking for solution for STE let us to choose the
parameterization for $R$ matrices as
\bea
\label{fff}
R_{xa}(u,v)=:e^{(a^1_{ij}(u,v)-\delta_{ij})c^+_ic_j}:,
\qquad R_{xyz}(u)=:e^{(a^3_{ij}(u)-\delta_{ij})c^+_ic_j}:, \nn \\
R_{yb}(u,v)=:e^{(a^2_{ij}(u,v)-\delta_{ij})c^+_ic_j}:,\qquad
R_{abz}(v)=:e^{(a^4_{ij}(v)-\delta_{ij})c^+_ic_j}:.
\ena
Then the (\ref{stede}) constraints can be rewritten as (we are
omitting three equations, which are identities, and we are surpressing the arguments
of the $a^i_{jk}$ functions)
\bea
{a^1_{21}a^3_{21}+a^1_{11}a^3_{23}a^4_{31}=a^4_{21}a^2_{21}}, &&
{a^1_{21}a^3_{12}+a^2_{11}a^3_{13}a^4_{32}=a^4_{12}a^1_{21}},\nn\\
{a^2_{21}a^3_{22}+a^2_{11}a^3_{23}a^4_{32}=a^4_{22}a^2_{21}}, &&
{a^1_{21}a^3_{11}+a^1_{11}a^3_{13}a^4_{31}=a^1_{21}a^4_{11}},\nn\\
{a^3_{13}a^4_{33}=a^3_{13}a^1_{22}+a^4_{13}a^3_{33}a^1_{21}}, &&
{a^3_{23}a^4_{33}=a^3_{23}a^2_{22}+a^4_{23}a^3_{33}a^2_{21}},\nn\\
{a^2_{12}a^4_{12}=a^3_{22}a^2_{12}+a^4_{23}a^3_{32}a^2_{11}}, &&
{a^1_{12}a^4_{21}=a^3_{21}a^2_{12}+a^4_{23}a^3_{31}a^2_{11}},\nn\\
{a^1_{22}a^3_{31}+a^1_{12}a^3_{33}a^4_{31}=a^4_{33}a^3_{31}},&&
{a^1_{22}a^3_{32}+a^2_{12}a^4_{32}a^3_{33}=a^4_{33}a^3_{32}},\nn\\
{a^2_{12}a^4_{12}=a^3_{12}a^1_{12}+a^4_{13}a^3_{32}a^1_{11}},&&
{a^2_{12}a^4_{22}=a^3_{22}a^2_{12}+a^4_{23}a^3_{32}a^2_{11}},\nn\\
{a^2_{22}a^3_{22}+a^2_{12}a^3_{23}a^4_{32}} &=&
{a^3_{22}a^2_{22}+a^4_{23}a^3_{32}a^2_{21}},
\qquad \qquad \quad \\
{a^1_{22}a^3_{11}+a^1_{12}a^3_{13}a^4_{31}} &=&
{a^3_{11}a^1_{22}+a^4_{13}a^3_{31}a^1_{21}},
\qquad \qquad \quad \nn\\
{a^1_{22}a^2_{21}+a^1_{12}a^3_{23}a^4_{31}} &=&
{a^3_{21}a^2_{22}+a^4_{23}a^3_{31}a^2_{21}},
\qquad \qquad \quad \nn\\
{a^2_{22}a^3_{12}+a^2_{12}a^3_{13}a^4_{32}} &=&
{ a^3_{12}a^1_{22}+a^4_{13}a^3_{32}a^1_{21}},
\qquad \qquad \quad \nn\\
{a^2_{21}a^3_{32}+a^2_{11}a^3_{33}a^4_{32}=a^4_{32}}, &&
{a^1_{21}a^3_{31}+a^1_{11}a^3_{33}a^4_{31}=a^4_{31}},\nn\\
{a^4_{23}=a^3_{23}a^2_{12}+a^4_{23}a^3_{33}a^2_{11}}, &&
{a^4_{13}=a^3_{13}a^1_{12}+a^4_{13}a^3_{33}a^1_{11}},\nn\\
{a^1_{11}a^4_{21}=a^4_{21}a^2_{11}},  && \nn
{a^2_{11}a^4_{12}=a^4_{12}a^1_{11}}.\label{par}
\ena

In the general case one has  the following constraints on the
$a_{ij}$ parameters
\bea
\label{equ}
a_{11}^{(1)}=a_{11}^{(2)}, &&
a_{22}^{(1)}=a_{22}^{(2)},\\\nn
\frac{a_{23}^{(3)}a_{32}^{(3)}}{a_{13}^{(3)}a_{31}^{(3)}}=
\frac{a_{23}^{(4)}a_{32}^{(4)}}{a_{13}^{(4)}a_{31}^{(4)}}, &&
\frac{a_{23}^{(3)}a_{32}^{(3)}}{a_{13}^{(3)}a_{31}^{(3)}}=
\frac{a_{22}^{(3)}-a_{22}^{(4)}}{a_{11}^{(3)}-a_{11}^{(4)}},\\\nn
\frac{{\chi_{13}^{12}}^{(3)}}{a_{23}^{(3)}}=
\frac{{\chi_{13}^{12}}^{(4)}}{a_{23}^{(4)}}, &&
\frac{{\chi_{23}^{21}}^{(3)}}{a_{13}^{(3)}}=
\frac{{\chi_{23}^{21}}^{(4)}}{a_{13}^{(4)}}, \\\nn
a_{11}^{(1)}=\frac{a_{11}^{(3)}-a_{11}^{(4)}}{a_{33}^{(4)}a_{11}^{(3)}
-{\chi_{13}^{13}}^{(4)}}, &&
a_{22}^{(2)}=\frac{a_{33}^{(4)}{\chi_{13}^{13}}^{(3)}-a_{33}^{(3)}{\chi_{13}^{13}}^{(4)}}
{a_{33}^{(4)}a_{11}^{(3)} -{\chi_{13}^{13}}^{(4)}},\\\nn
a_{21}^{(1)}=\frac{a_{31}^{(3)}a_{13}^{(4)}}{a_{33}^{(4)}a_{11}^{(3)}
-{\chi_{13}^{13}}^{(4)}}, &&
a_{12}^{(1)}=\frac{a_{31}^{(4)}a_{13}^{(3)}}{a_{33}^{(4)}a_{11}^{(3)}
-{\chi_{13}^{13}}^{(4)}},\\\nn
a_{12}^{(2)}=\frac{a_{23}^{(3)}a_{31}^{(4)}a_{13}^{(4)}}{a_{23}^{(4)}
(a_{33}^{(4)}a_{11}^{(3)}-{\chi_{13}^{13}}^{(4)})}, &&
a_{21}^{(2)}=\frac{a_{23}^{(4)}a_{31}^{(3)}a_{13}^{(3)}}{a_{23}^{(3)}
(a_{33}^{(4)}a_{11}^{(3)}-{\chi_{13}^{13}}^{(4)})},\\\nn
{\chi_{ij}^{kr}}^{(\alpha)} =
a_{ki}^{(\alpha)}a_{rj}^{(\alpha)} &-& a_{kj}^{(\alpha)}a_{ri}^{(\alpha)},\qquad
\alpha=3,4.
\ena
We will pay special attention to the homogeneous case
\bea
a_{ij}^{(r)}=a_{ji}^{(r)}, \quad r=1,2,3,4,\quad
a_{11}^{(r)}=a_{22}^{(r)},\quad r=1,2, \quad
a_{11}^{(\alpha)}=a_{22}^{(\alpha)}=a_{33}^{(\alpha)},\quad
\alpha=3,4,
\ena
where the above equations can be  converting into the following equations:
\bea
\frac{{\chi_{13}^{13}}^{(r)}+1}{a_{33}^{(r)}}=
\frac{{\chi_{12}^{12}}^{(\alpha)}+1}{a_{11}^{(\alpha)}}\equiv
\mb{const}=\Delta, \qquad \alpha=1,2,\,r=3,4,\\
a_{ij}^{(1)}=a_{ij}^{(2)}\quad a_{13}^{(r)}=a_{23}^{(r)},\quad
a_{12}^{(r)}=a_{11}^{(r)}+\bar{\Delta},\quad
r=3,4,\,\bar{\Delta}=\mb{const}. \ena The solution of these
equations for  $R_2$  can be parameterized via trigonometric
functions \bea a_{11}=a_{22}=\sin (u)/\cos(u), \quad
a_{12}=a_{21}=1/\cos(u),\quad \Delta=0. \ena This is the
parameterization of the XX-model. For for the $R_3$ we then obtain
\bea \label{3XX}
&&a_{12}=a_{21}=a_{11}=a_{22}=a_{33}=\sin(u)/\cos(u), \nn \\
&&a_{13}=a_{31}=a_{23}=a_{32}=1/\cos(u), \qquad \qquad \qquad
\bar{\Delta}=0, \ena which is an extension of the XX R-matrix to
three dimensional case. Then, by using the technique of coherent
states developed in \cite{coh,AS} one can find a fermionic
expressions of the transfer matrix $\tau(u)$  and the partition
function  $Z=\prod^{N}\tau(u)$: \bea \tau(u)&=&\int
D\bar{\psi}D\psi
:\exp\{\sum_{i,j}^{\frac{N}{2},\frac{N}{2}}\left(a_{12}\bar{\psi}_{x,i,j}\psi_{y,i-1,j-1}
+a_{21}\bar{\psi}_{y,i,j}
\psi_{x,i-1,j-1}+a_{11}\bar{\psi}_{x,i,j}\psi_{x,i-2,j}\right. \nn \\
&+&\left. a_{22}\bar{\psi}_{y,i,j}\psi_{y,i,j-2}-
+a_{13}\bar{\psi}_{x,i,j}c_{z,i-1,j}+a_{31}\bar{c}_{z,i,j}\psi_{x,i-1,j}+
a_{23}\bar{\psi}_{y,i,j}c_{z,i,j-1}\right. \nn \\
&+&\left. a_{32}\bar{c}_{z,i,j}\psi_{y,i,j-1}+
(a_{33}-1)\bar{c}_{z,i,j}c_{z,i,j}-\bar{\psi}_{x,i,j}\psi_{x,i,j}-\bar{\psi}_{y,i,j}\psi_{y,i,j}\right)\}:.
\ena After Fourier transformation we will have \bea \label{f}
\tau(u)&=&\int D\bar{\psi}D\psi
\prod_{p,q}^{\frac{N}{2},\frac{N}{2}}:\exp\{
\left(a_{12}e^{i(p+q)}\bar{\psi}_{x,p,q}\psi_{y,p,q}+a_{21}e^{i(p+q)}\bar{\psi}_{y,p,q}
\psi_{x,p,q}+a_{11}e^{2ip}\bar{\psi}_{x,p,q}\psi_{x,p,q}\right. \nn \\
&+&\left.a_{22}e^{2iq}\bar{\psi}_{y,p,q}\psi_{y,p,q}+
a_{13}e^{ip}\bar{\psi}_{x,p,q}c_{z,p,q}+a_{31}e^{ip}\bar{c}_{z,p,q}\psi_{x,p,q}+
a_{23}e^{iq}\bar{\psi}_{y,p,q}c_{z,p,q} \right. \nn \\
&+&\left.a_{32}e^{iq}\bar{c}_{z,p,q}\psi_{y,p,q}+(a_{33}-1)\bar{c}_{z,p,q}c_{z,p,q}-
\bar{\psi}_{x,p,q}\psi_{x,p,q}-\bar{\psi}_{y,p,q}\psi_{y,p,q}\right)\}= \nn \\
&=&\prod_{p,q}^{\frac{N}{2},\frac{N}{2}}a(p,q):\exp\{A_{p,q}\bar{c}_{z,p,q}c_{z,p,q}\}:.
\ena Here the normal ordering notion :: refers to the  $\bar
c_{z}$ and $ c_{z}$ operators and \bea \label{f2}
a(p,q)&=&1+(a_{11}a_{22}-a_{12}a_{21})e^{2i
(p+q)}-(a_{11}e^{2i p}+a_{22}e^{2 i q}), \qquad  \nn \\
A_{p,q}&=&a_{33}-1-\frac{1}{a(p,q)}
\left((a_{22}e^{2iq}-1)a_{13}a_{31}e^{2i p}+(a_{11}
e^{2i p}-1)a_{32}a_{23}e^{2iq}\right. \nn \\
&-&\left.a_{12}a_{31}a_{23}e^{2i(p+q)}-a_{21}a_{13}a_{32}e^{2i(p+q)}\right).
\ena
The solution corresponding to (\ref{3XX}) reads
\bea
a(p,q)&=&1-\sin(u)/\cos(u)(e^{2ip}+e^{2iq}), \\
A_{p,q}&=&\sin(u)/\cos(u)-1+\frac{e^{2ip}+e^{2iq}}{\cos(u)^2(1-\sin(u)/\cos(u)(e^{2ip}+e^{2iq}))}. \nn
\ena

At the point $u=0$ we have
\bea
a(p,q)&=&1-u(e^{2ip}+e^{2iq}), \nn \\
A_{p,q}&=&u-1+(e^{2ip}+e^{2iq})(1+u(e^{2ip}+e^{2iq})),
\ena
and the transfer matrix $\tau(u)$ takes the following form for
small values of the spectral parameter $u$
\bea \label{f3}
\tau&=&\prod_{p,q}^{\frac{N}{2},\frac{N}{2}}
(a(p,q):\exp\{A_{p,q}\bar{c}_{z,p,q}c_{z,p,q}\}:)=\nn \\
&=&\prod_{p,q}^{\frac{N}{2},\frac{N}{2}}
a(p,q):e^{(u-1+(e^{2ip}+e^{2iq})(1+u(e^{2ip}+e^{2iq})))\bar{c}_{z,p,q}c_{z,p,q}}:=\nn \\
&=&\prod_{p,q}^{\frac{N}{2},\frac{N}{2}}
e^{P(p,q)\bar{c}_{z,p,q}c_{z,p,q}+u\,\varepsilon(p,q)\bar{c}_{z,p,q}c_{z,p,q}},
\ena where \bea e^{P(p,q)}=e^{2ip}+e^{2iq},\qquad
\varepsilon(p,q)=2\cosh{P(p,q)}. \qquad \ena
When  $ip\rightarrow-\infty$ ( or
$iq\rightarrow-\infty$) this transfer matrix decouples into the
product of N independent XX model's transfer matrices defined on a
1D chain.


In the case $a_{ii}=0$ there are another solutions for the
fermionic  STE equations (\ref{par}). They are defined by three
multiplicative relations for the parameters of three-particle
R-matrices \bea a^k_{21}a^k_{12}=\beta,\quad a^k_{13}=\alpha
a^k_{12}a^k_{23},\quad a^k_{31}    =\gamma a^k_{32}a^k_{21},\quad
k=3,4, \ena where the constants $\alpha,\beta,\gamma$ can be
considered as model parameters.  The relations can be reformulated
as \bea a_{21}(u)a_{12}(u)=\beta, \quad a_{13}(u)=\alpha
a_{12}(u)a_{23}(u), \quad a_{31}(u)=\gamma a_{32}(u)a_{21}(u).\ena

The non zero matrix elements of corresponding $R_{ijk}$ operators
in this case are the following (with above relations) \be \ba{lll}
R_{000}^{000}=1,
&R_{100}^{001}=a_{12},&R_{001}^{100}=a_{21},\\
R_{011}^{011}=a_{32}a_{23},&R_{100}^{010}=a_{13},&R_{010}^{100}=a_{31},\\
R_{110}^{110}=a_{13}a_{31},
&R_{011}^{110}=-a_{23}a_{31},&R_{110}^{011}=-a_{13}a_{32},\\
R_{101}^{101}=-a_{12}a_{21},
&R_{101}^{011}=a_{23}a_{12},&R_{110}^{101}=-a_{31}a_{12},\\
R_{011}^{101}=-a_{32}a_{21},&R_{010}^{001}=a_{32},&
R_{001}^{010}=-a_{23},\\
R_{101}^{110}=-a_{13}a_{21},
&R_{111}^{111}=a_{13}a_{32}a_{21}+a_{23}a_{31}a_{12}.&
 \ea \label{R3coh}
 \ee
While  the non vanishing matrix elements of two $(R^{1,2})_{ij}$
intertwiners as solutions of the semi-tetrahedron equations have
the following expressions in terms of the $a_{ij}$ functions of
$R_{123}$'s
 \be
 \ba{cc}(R^1(u,v))_{00}^{00}=a_{32}(u)/a_{32}(v),&(R^1(u,v))_{11}^{11}=a_{21}(v)/a_{21}(u),\\
(R^1(u,v))_{01}^{10}=a_{32}(u)a_{21}(v)/a_{32}(v)a_{21}(u),&(R^1(u,v))_{10}^{01}=1,\\
(R^2(u,v))_{00}^{00}=a_{12}(u)/a_{12}(v),&(R^2(u,v))_{11}^{11}=a_{23}(v)/a_{23}(u),\\
(R^2(u,v))_{01}^{10}=a_{12}(u)a_{23}(v)/a_{12}(v)a_{23}(u),&(R^2(u,v))_{10}^{01}=1.
 \ea \label{R2coh}
 \ee

As in two dimensional case we can extract from the commuting transfer
matrices exact solvable quantum models (in our case on a two-dimensional lattice
rather than on one-dimensional lattice),
identifying first the logarithmic derivative of
the transfer matrix with the  Hamiltonian:
\bea \nn \ln \tau(u)=\ln
\left(tr_{x_i,y_i}\prod_{i,j}R_{x_i,y_i,z_{i,j}}(u)\right)=\\\nn
=\ln \tau(u_0)+u (\tau(u_0))^{-1}d\tau(u)/du|_{u=u_0}+ ... . \ena

By inserting the representation (\ref{a})  for $R_{x_i,y_i,z_{i,j}}(u)$
in this equation and taking the trace over the auxiliary spaces,
due to the gaussian form of the exponents in
R-operators, $\tau$ will have the similar form (in the following
we shall omit the $z_{ij}$ notation, keeping only {ij})
\bea
\label{Ham}
\tau &=& (1-\beta^{N})^{N}:e^{\frac{1}{(1-\beta^N)^{N}}H(c^+,c)}:, \nn \\
H(c^+,c) &=& \sum_{i,j}\sum_r A_r
c^+_{i,j}c_{i+r,j-r}+\sum_{i,j}\sum_r B_r
c^+_{i,j}c_{i+r+1,j-r}+\sum_{i,j}\sum_r C_r
c^+_{i,j}c_{i+r,j-r-1}, \nn \\
A_r&=&(\beta)^{r-1}(\frac{1}{\alpha}+\frac{1}{\gamma})a_{13}a_{31},\qquad
B_r=(\beta)^{r}a_{13}a_{31},\qquad C_r=(\beta)^{r}a_{32}a_{23}.
 \ena

From (22) it follows that
$$a_{23}a_{32}=a_{13}a_{31}/(\alpha\beta\gamma).$$
If $a_{13}a_{31}$
is small it is possible to expand $\tau$ in terms of
the $a_{13}a_{31}$ variables and  $H$ in the expression above
manifests itself as a non-local Hamiltonian. From the structure of
$H$ it follows that although Hamiltonian is not local, there is
localization around the path $\left({i,j}\rightarrow {i+1,j+1}
\rightarrow ...\rightarrow {i+r,j-r}...\right)$. In momentum space
this becomes more apparent
\bea
\mt{H}(c^+_{p_x,p_y},c_{p_x,p_y})&=&\sum_{p_x,p_y
=p_x+p_0}\frac{1}{\beta}(\frac{1}{\alpha}+\frac{1}{\gamma} +\beta
e^{i p_x}+\frac{1}{\alpha\gamma}e^{-i p_x +i
p_0})c^+_{p_x,p_x+p_0},c_{p_x,p_x+p_0},\nn \\
&&  \qquad \qquad \qquad \qquad  \qquad \qquad \qquad \qquad p_0 = i\ln{\beta}.
 \ena

In the case where
\be \alpha\gamma=(\beta)^2, \qquad \qquad \qquad \frac{\alpha}{\gamma}=e^{p_1}
\ee
the Hamiltonian is  real and can be written as
\bea
\label{H}
\mt{H}(c^+_{p_x,p_y},c_{p_x,p_y})&=&2\sum_{p}(\cosh{p_1}+\cos{p})c^+_{p,p+p_0},c_{p,p+p_0},\nn \\
p_0=i\ln{\alpha\gamma}/2,  &&  p_1=\ln{\frac{\alpha}{\gamma}}.
 \ena
In the limits $\beta\rightarrow 0,\infty$ Hamiltonian (\ref{Ham})
becomes completely local.

The expression (\ref{H})  allows us to represent the
transfer matrix as $\tau=\exp{\mt{H}}$, and thus to find the exact
discrete time Hamiltonian not only for  small values of $a_{12}a_{21}$.
The Hamiltonian follows directly from:
$$\tau=A:e^{H}:=A\prod_{k}:e^{\varepsilon(k)c_k^+c_k}=
Ae^{\epsilon(k)c_k^+c_k}, \qquad
\epsilon(k)=\ln{(1+\varepsilon(k))}.
$$

As one can  see  from the expression (\ref{H})  rotational
symmetry is broken in the above model. In order to  recover it we
can suggest another integrable model , where the transfer matrix
is constructed with the help of the same solutions $R_3$ and $R_2$
(satisfying (\ref{R3coh}) and (\ref{R2coh})) of the local  STE
equations (\ref{sted}), as in the previous section, but with some
chess like disposition of the $R_3$ and rotated $R_3$ matrices
\bea \bar{\tau}(u)=tr \bar{T}(u)
=\sum_{\{x,y\}}\prod_{j}\left(\prod_{i}R_{x_{2i},y_{2j},z_{2i,2j}}(u)
R_{x_{2i+1},y_{2j},z_{2i+1,2j}}^{\tau_1}(u)\right. \nn \\
\left.\prod_{i}R_{x_{2i},y_{2j+1},z_{2i,2j+1}}^{\tau_2}(u)
R_{x_{2i+1},y_{2j+1},z_{2i+1,2j+1}}^{\tau_3}(u)\right),
\label{tau1} \ena where $\tau_{i}$, $i=1,2,3$, refer to the
rotated matrices \bea
{R^{\tau_1}}_{ijk}^{i'j'k'}={R}_{ij'k}^{i'jk'},\nn\\
{R^{\tau_2}}_{ijk}^{i'j'k'}={R}_{i'jk}^{ij'k'}, \label{tra}\\
{R^{\tau_3}}_{ijk}^{i'j'k'}={R}_{i'j'k}^{ijk}.\nn
\ena
In  this case the multiplicative intertwiner matrix,
necessary for the commutativity of two transfer matrices with
different spectral parameters:
\bea
\bar{R}_{a,b;x,y}(u,v)\bar{T}(u)\bar{T}(v)=
\bar{T}(v)\bar{T}(u)\bar{R}_{a,b;x,y}(u,v) , \label{intco}
\ena
is modified a little compared to (\ref{int})
\be
\bar{R}_{a,b;x,y}(u,v)=\prod_{j}R_{y_{2j},b_{2j}}(u,v)R_{y_{2j+1},b_{2j+1}}^{\tau
-1}(u,v)
\prod_{i}R_{x_{2i},a_{2i}}(u,v)R_{x_{2i+1},a_{2i+1}}^{\tau-1}(u,v).
\ee
Here $\tau$ denotes
\be
{R_{ab}^{cd}}^{\tau}=R_{cd}^{ab}
\label{tau}
\ee

The commutativity can be verified by putting the right hand sides of (\ref{tau1})
and (\ref{tau}) into the equation (\ref{intco}) and observing
 that it is fulfilled if  four local equations are satisfied
(as it was done in (\ref{com}) and (\ref{sem})).  Here we are omitting all
arguments of the operators:
\bea
R_{yb}R_{xa}R_{xyz}R_{abz}&=&R_{abz}R_{xyz}R_{xa}R_{yb},\label{stedn}\\
({R_{xa}}^{\tau_x\tau_a})^{-1}R_{yb}{R_{xyz}}^{\tau_x}{R_{abz}}^{\tau_a}
&=&{R_{abz}}^{\tau_a}{R_{xyz}}^{\tau_x}R_{yb}({R_{xa}}^{\tau_x\tau_a})^{-1},\label{stedn1}\\
({R_{yb}}^{\tau_y\tau_b})^{-1}R_{xa}{R_{xyz}}^{\tau_y}{R_{abz}}^{\tau_b}
&=&{R_{abz}}^{\tau_b}{R_{xyz}}^{\tau_y}R_{xa}({R_{yb}}^{\tau_y\tau_b})^{-1}, \label{stedn2}\\
({R_{yb}}^{\tau_y\tau_b})^{-1}
({R_{xa}}^{\tau_x\tau_a})^{-1}{R_{xyz}}^{\tau_x\tau_y}{R_{abz}}^{\tau_a\tau_b}
&=&{R_{abz}}^{\tau_a\tau_b}{R_{xyz}}^{\tau_x\tau_y}({R_{xa}}^{\tau_x\tau_a})^{-1}
({R_{yb}}^{\tau_y\tau_b})^{-1}. \label{stedn3}
 \ena
 By $\tau_x,\tau_y,\tau_a,\tau_b$ we mean the rotations or
 transpositions  defined in (\ref{tra}) and (\ref{tau})
 \bea
{R_{xa}}^{\tau_x\tau_a}={R_{xa}}^{\tau},&\qquad
&{R_{yb}}^{\tau_y\tau_b}={R_{yb}}^{\tau},\nn\\
{R_{xyz}}^{\tau_x}={R_{xyz}}^{\tau_2},&\qquad&
{R_{abz}}^{\tau_a}={R_{abz}}^{\tau_2},\nn\\
{R_{xyz}}^{\tau_y}={R_{xyz}}^{\tau_1},&\qquad&
{R_{abz}}^{\tau_b}={R_{abz}}^{\tau_1},\label{tra1}\\
{R_{xyz}}^{\tau_x\tau_y}={R_{xyz}}^{\tau_3},&\qquad&
{R_{abz}}^{\tau_a\tau_b}={R_{abz}}^{\tau_3}.\nn
 \ena

The solutions of the first equation (\ref{stedn}) (which are the
STE defined in (\ref{sted})) are also solutions for other three
equations.  It can be shown for (\ref{stedn1}) by
applying corresponding $\tau$ operation on the standard
 STE equations as  done below:
\bea
(R_{yb}R_{xa}R_{xyz}R_{abz})^{\tau_x \tau_a}&=&(R_{abz}R_{xyz}R_{xa}R_{yb})^{\tau_x \tau_a},\\
R_{yb}{R_{xyz}}^{\tau_x}{R_{abz}}^{\tau_a}{R_{xa}}^{\tau_x\tau_a}&=
&{R_{xa}}^{\tau_x\tau_a}{R_{abz}}^{\tau_a}{R_{xyz}}^{\tau_x}R_{yb},
\ena
Two other equations, (\ref{stedn2}) and (\ref{stedn3}), can be derived analogously.

\begin{figure}[t]
\label{man}
\unitlength=25pt
\begin{picture}(100,8)(-3,2)
\multiput(0,3)(0,1){7}{\line(1,0){11}}
\multiput(1,2)(1,0){10}{\line(0,1){8}}
\put(4,7){{\linethickness{1pt}\vector(1,0){1}}}
\put(4,6){{\linethickness{1pt}\vector(0,1){1}}}
\put(5,7){{\linethickness{1pt}\vector(0,-1){1}}}
\put(5,6){{\linethickness{1pt}\vector(-1,0){1}}}
\put(4,6){{\linethickness{1pt}\vector(-1,0){1}}}
\put(3,6){{\linethickness{1pt}\vector(0,-1){1}}}
\put(3,5){{\linethickness{1pt}\vector(1,0){1}}}
\put(4,5){{\linethickness{1pt}\vector(0,1){1}}}
\put(-0.5,6){$2j$}\put(-0.5,7){$2j+1$} \put(3.8,1.5){$2i$}
\put(4.8,1.5){$2i+1$} \put(4,6){\circle*{0.3}}
\end{picture}
\caption{Manhattan like structure on 2d square lattice}
\end{figure}
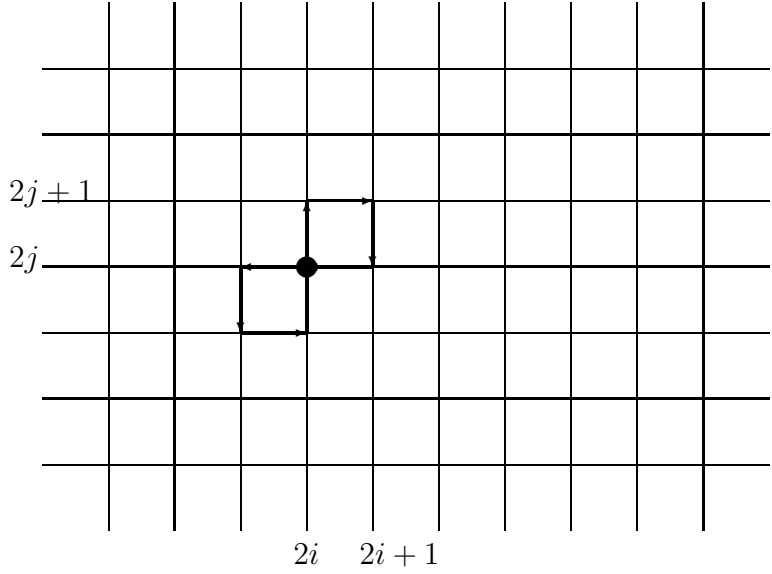

 The corresponding 2d quantum Hamiltonian in the fermionic
representation for this model, unlike the previous one, has only
nearest-neighbor and next to nearest-neighbor interactions, and
exhibits a Manhattan like structure on the 2d square lattice shown in  Fig. \ref{man}.
\bea
\bar{\tau}&=&(1-\beta^2)^{\frac{N \times
N}{2}}:e^{\frac{1}{1-\beta^2}\bar{H}(c^+,c)}:  \; ,  \\
\bar{H}(c^+,c)&=&a_{31}a_{13}\sum_{i,j}(c^+_{2i+1,2j}c_{2i,2j}+
1/\alpha c^+_{2i+1,2j-1}c_{2i,2j}+\beta
c^+_{2i,2j-1}c_{2i,2j}+\beta/\alpha
c^+_{2i,2j}c_{2i,2j}) \nn \\
&+& a_{23}a_{32}\sum_{i,j}(c^+_{2i,2j+1}c_{2i,2j}+ \beta\alpha
c^+_{2i-1,2j+1}c_{2i,2j}+\beta
c^+_{2i-1,2j}c_{2i,2j}+\beta^2\alpha c^+_{2i,2j}c_{2i,2j})+... .\nn
\label{Ham1}
 \ena

The $\bar{H}(c^+,c)$ consists of mass terms and hopping terms only
along the elementary circuits, described by the arrows drawn on
the links of the square lattice in the Fig.\ref{man} (if continued for
all the links, they form a square Manhattan lattice). In
(\ref{Ham1}) we find  terms with $c_{2i,2j}$
operators. They correspond to the circulations around two
elementary circuits with clockwise and counterclockwise
orientations, connected by the $2i,2j$ vertex. The other three
terms with the annihilation operators on the
 even-odd, odd-odd, even-even sites can be obtained from that
expression, following to the directions of the arrows, outgoing
from that site and writing the same horizontal, diagonal and
vertical hopping parameters (and also the mass terms) as in the
first term.

 In the momentum space
 \be
 \ba{ll}
c_{2j,2k}=\frac{1}{N}\sum_{p,q=1}^{N/2}e^{-2\pi\frac{2jp+2kq}{N}}c_{1p,q},&
c^+_{2j,2k}=\frac{1}{N}\sum_{p,q=1}^{N/2}e^{2\pi\frac{2jp+2kq}{N}}c^+_{1p,q},\nn\\
c_{2j+1,2k}=\frac{1}{N}\sum_{p,q=1}^{N/2}e^{-2\pi\frac{(2j+1)p+2kq}{N}}c_{2p,q},&
c^+_{2j+1,2k}=\frac{1}{N}\sum_{p,q=1}^{N/2}e^{2\pi\frac{(2j+1)p+2kq}{N}}c^+_{2p,q},\nn\\
c_{2j+1,2k+1}=\frac{1}{N}\sum_{p,q=1}^{N/2}e^{-2\pi\frac{(2j+1)p+(2k+1)q}{N}}c_{3p,q},&
c^+_{2j+1,2k+1}=\frac{1}{N}\sum_{p,q=1}^{N/2}e^{2\pi\frac{(2j+1)p+(2k+1)q}{N}}c^+_{3p,q},\nn\\
c_{2j,2k+1}=\frac{1}{N}\sum_{p,q=1}^{N/2}e^{-2\pi\frac{2jp+(2k+1)q}{N}}c_{4p,q},&
c^+_{2j,2k+1}=\frac{1}{N}\sum_{p,q=1}^{N/2}e^{2\pi\frac{2jp+(2k+1)q}{N}}c^+_{4p,q}
\ea
\ee
the Hamiltonian  (\ref{Ham1}) will take a more compact form
\bea\nn \bar{H}(p,q)&=&
c^+_{1p,q}c_{2p,q}(a_{31}a_{13}\beta e^{-ip}+a_{23}a_{32}
e^{ip})+ c^+_{2p,q}c_{1p,q}(a_{31}a_{13} e^{ip}+a_{23}a_{32}\beta
e^{-ip})\nn \\
&+& c^+_{1p,q}c_{3p,q}(a_{31}a_{13}\frac{1}{\alpha}
e^{-ip+iq}+a_{23}a_{32}\beta\alpha e^{ip-iq})+
c^+_{3p,q}c_{1p,q}(a_{31}a_{13}\frac{1}{\alpha}
e^{ip-iq}+a_{23}a_{32}\beta\alpha e^{-ip+iq})\nn \\
&+&
c^+_{1p,q}c_{4p,q}(a_{31}a_{13}\beta e^{iq}+a_{23}a_{32}\beta
e^{-iq})+ c^+_{4p,q}c_{1p,q}(a_{31}a_{13} \beta
e^{-iq}+a_{23}a_{32}e^{iq})\nn \\
&+&
c^+_{2p,q}c_{3p,q}(a_{31}a_{13}\beta e^{iq}+a_{23}a_{32} e^{-iq})+
c^+_{3p,q}c_{2p,q}(a_{31}a_{13} e^{-iq}+a_{23}a_{32}\beta
e^{iq})\nn\\
&+& c^+_{2p,q}c_{4p,q}(a_{31}a_{13}\beta\alpha
e^{ip+iq}+a_{23}a_{32}\frac{1}{\alpha} e^{-ip-iq})+
c^+_{4p,q}c_{2p,q}(a_{31}a_{13}\beta\alpha
e^{-ip-iq}+a_{23}a_{32}\frac{1}{\alpha} e^{ip+iq})\nn\\
&+&
c^+_{3p,q}c_{4p,q}(a_{31}a_{13}\beta e^{ip}+a_{23}a_{32}e^{-ip})+
c^+_{4p,q}c_{3p,q}(a_{31}a_{13} e^{-ip}+a_{23}a_{32}\beta
e^{ip})\nn \\
&+&
 (c^+_{1p,q}c_{1p,q}+c^+_{3p,q}c_{3p,q})(a_{31}a_{13}\frac{1}{\alpha}
+a_{23}a_{32}
\beta^2\alpha) \\
&+&(c^+_{2p,q}c_{2p,q}+c^+_{4p,q}c_{4p,q})(a_{31}a_{13}
\beta^2\alpha+a_{23}a_{32}\frac{1}{\alpha}). \nn
\ena

For the simple choice of the parameters
$a_{31}a_{13}=a_{23}a_{32}=a^2$, $\alpha=1$, $\beta=1$, the
Hamiltonian  $\bar{H}(p,q)$ acquires simple matrix form
\be
 2a^2 \left( \ba{llll} 1&\cos{p}&\cos{(p-q)}&\cos{q}\\
\cos{p}&1&\cos{q}&\cos{(p+q)}\\
\cos{(p-q)}&\cos{q}&1&\cos{p}\\
\cos{q}&\cos{(p+q)}&\cos{p}&1 \ea \right), \ee two nonzero
eigenvalues of which gives us the spectrum
 \be e_{\pm}=8 a^2
(1\pm\cos{p}\cos{q}).
\ee

\subsection*{Acknowledgment}
J.A. and A.S partially were  supported by ``MaPhySto'', the Center of
Mathematical Physics and Stochastics, financed by the National
Danish Research Foundation as well as by INTAS grant
03-51-5460. Sh.Kh. acknowledges INTAS grant 03-51-5460
for support.

\subsection*{Appendix:  ZTE equations for  free fermionic
representation (\ref{fff}) of R-operators}

In this appendix we present the independent set of ZTE equations
(\ref{tetr}) in case when the three state R-operators admit the
$:e^{(a_{ij}-\delta_{ij})c^+_{i}c_{j}}:$ free-fermionic
realization.  For the operators (\ref{matrixR}) the equations
(\ref{tetr}), written in matrix form, are:
 \bea
\sum_{j_\alpha}{R_1}_{i_c i_x i_a}^{j_c j_x j_a}{R_2}_{j_c i_y
i_b}^{k_c j_y j_b}{R_3}_{j_x j_y i_z}^{k_x k_y j_z} {R_4}_{j_a j_b
j_z}^{k_a k_b
k_z}(v)(-1)^{p(j_x)p(j_b)+p(k_y)p(k_a)+p(i_z)p(k_c)}=\nn
\\
\sum_{j_\alpha}{R_4}_{i_a i_b i_z}^{j_a j_b j_z}{R_3}_{i_x i_y
j_z}^{j_x j_y k_z}{R_2}_{i_c j_y j_b}^{j_c k_y k_z}{R_1}_{j_c j_x
j_a}^{k_c k_x
k_a}(-1)^{p(j_x)p(j_b)+p(i_y)p(i_a)+p(i_c)p(k_z)}.\label{ztede}
\ena In the equations we have omitted spectral parameters and we
have distinguished the elements of different matrices by labels
$1,2,3,4$. In terms of $a^k_{ij}$ elements the non-identity
equations are \be
 \ba{c}
a^1_{12}a^2_{21}a^3_{12}=
a^3_{21}a^2_{12}a^1_{21},\\
a^1_{22}a^3_{31}a^4_{13}+a^1_{32}a^4_{11}+a^1_{12}a^2_{31}a^4_{12}
+a^1_{12}a^2_{21}a^3_{32}a^4_{13}=
a^3_{21}a^2_{12}a^1_{31}+a^3_{11}a^1_{32},\\
a^1_{22}a^3_{21}+a^1_{12}a^2_{21}a^3_{22}=a^3_{21}a^2_{22},\\
a^1_{22}a^3_{31}a^4_{23}+a^1_{32}a^4_{21}+a^1_{12}a^2_{31}a^4_{22}
+a^1_{12}a^2_{21}a^3_{32}a^4_{23}=
a^3_{21}a^2_{32},\\
a^1_{22}a^3_{31}a^4_{33}+a^1_{32}a^4_{31}+a^1_{12}a^2_{31}a^4_{32}
+a^1_{12}a^2_{21}a^3_{32}a^4_{33}=a^3_{31},\\
a^1_{12}a^2_{11}=a^3_{11}a^1_{12}+a^3_{21}a^2_{12}a^1_{11},\\
a^1_{23}a^3_{11}+a^1_{13}a^2_{21}a^3_{12}
=a^4_{21}a^2_{13}a^1_{21}+a^4_{11}a^1_{23}+
a^4_{31}a^3_{23}a^2_{12}a^1_{21}+a^4_{31}a^3_{13}a^1_{22},\\
a^1_{13}a^2_{21}a^3_{32}a^4_{13}+a^1_{13}a^2_{31}a^4_{12}
+a^1_{23}a^3_{31}a^4_{13}=a^4_{21}a^2_{13}a^1_{31}+
a^4_{31}a^3_{23}a^2_{12}a^1_{31}+a^4_{31}a^3_{13}a^1_{32},\\
a^1_{23}a^3_{21}+a^1_{13}a^2_{21}a^3_{22}=
a^4_{31}a^3_{23}a^2_{22}+a^4_{21}a^2_{23},\\
a^1_{33}a^4_{21}+a^1_{13}a^2_{21}a^3_{32}a^4_{23}+a^1_{13}a^2_{31}a^4_{22}
+a^1_{23}a^3_{31}a^4_{23}=a^4_{31}a^3_{23}a^2_{32}+a^4_{21}a^2_{33},\\
a^1_{33}a^4_{31}+a^1_{13}a^2_{21}a^3_{32}a^4_{33}+a^1_{13}a^2_{31}a^4_{32}
+a^1_{23}a^3_{31}a^4_{33}=a^4_{31}a^3_{33},
 \ea \ee \be\nn \ba{c}
a^1_{13}a^2_{11}=a^4_{21}a^2_{13}a^1_{11}+a^4_{31}a^3_{23}a^2_{12}a^1_{11}
+a^4_{31}a^3_{13}a^1_{12}+a^4_{11}a^1_{13},\\
a^2_{22}a^3_{12}=a^3_{12}a^1_{22}+a^3_{22}a^2_{12}a^1_{21},\\
a^2_{22}a^3_{32}a^4_{13}+a^2_{32}a^4_{12}
=a^3_{22}a^2_{12}a^1_{31}+a^3_{12}a^1_{32},\\
a^2_{22}a^3_{32}a^4_{23}+a^2_{32}a^4_{22}=a^3_{22}a^2_{32},\\
a^2_{22}a^3_{32}a^4_{33}+a^2_{32}a^4_{32}=a^3_{32},\\
a^2_{12}=a^3_{12}a^1_{12}+a^3_{22}a^2_{12}a^1_{11},\\
a^2_{23}a^3_{12}=a^4_{22}a^2_{13}a^1_{21}+a^4_{12}a^1_{23}+
a^4_{32}a^3_{23}a^2_{12}a^1_{21}+a^4_{32}a^3_{13}a^1_{22},\\
a^2_{23}a^3_{32}a^4_{13}+a^2_{33}a^4_{12}=a^4_{22}a^2_{13}a^1_{31}+a^4_{12}a^1_{33}+
a^4_{32}a^3_{23}a^2_{12}a^1_{31}+a^4_{32}a^3_{13}a^1_{32},\\
a^2_{23}a^3_{22}=a^4_{22}a^2_{23}+a^4_{32}a^3_{23}a^2_{22},\\
a^2_{23}a^3_{32}a^4_{23}=a^4_{32}a^3_{23}a^2_{32},\\
a^2_{23}a^3_{32}a^4_{33}+a^2_{33}a^4_{32}=a^4_{32}a^3_{33},\\
a^2_{13}=a^4_{22}a^2_{13}a^1_{11}+a^4_{12}a^1_{13}+
a^4_{32}a^3_{23}a^2_{12}a^1_{11}+a^4_{32}a^3_{13}a^1_{12},\\
a^3_{13}=a^4_{23}a^2_{13}a^1_{21}+a^4_{13}a^1_{23}+
a^4_{33}a^3_{23}a^2_{12}a^1_{21}+a^4_{33}a^3_{13}a^1_{22},\\
a^3_{33}a^4_{13}=a^4_{23}a^2_{13}a^1_{31}+a^4_{13}a^1_{33}+
a^4_{33}a^3_{23}a^2_{12}a^1_{31}+a^4_{33}a^3_{13}a^1_{32},\\
a^3_{23}=a^4_{33}a^3_{23}a^2_{22}+a^4_{23}a^2_{23},\\
a^3_{33}a^4_{23}=a^4_{33}a^3_{23}a^2_{32}+a^4_{23}a^2_{33},\\
0=a^4_{23}a^2_{13}a^1_{11}+a^4_{13}a^1_{13}+
a^4_{33}a^3_{23}a^2_{12}a^1_{11}+a^4_{33}a^3_{13}a^1_{12},\\
a^1_{21}a^3_{11}+a^1_{11}a^2_{21}a^3_{12}=a^2_{11}a^1_{21},\\
a^1_{31}a^4_{11}+a^1_{11}a^2_{21}a^3_{32}a^4_{13}+a^1_{11}a^2_{31}a^4_{12}
+a^1_{21}a^3_{31}a^4_{13}=a^2_{11}a^1_{31},\\
a^1_{21}a^3_{21}+a^1_{11}a^2_{21}a^3_{22}=a^2_{21},\\
a^1_{31}a^4_{21}+a^1_{11}a^2_{21}a^3_{32}a^4_{23}+a^1_{11}a^2_{31}a^4_{22}
+a^1_{21}a^3_{31}a^4_{23}=a^2_{31},\\
a^1_{31}a^4_{31}+a^1_{11}a^2_{21}a^3_{32}a^4_{33}+a^1_{11}a^2_{31}a^4_{32}
+a^1_{21}a^3_{31}a^4_{33}=0. \ea \ee

\end{document}